\documentclass[12pt,twocolumn]{article}
\usepackage{amsmath,amsfonts,amssymb}
\usepackage{graphicx}
\usepackage{caption}
\usepackage[margin=0.7in]{geometry}
\usepackage[utf8]{inputenc}
\usepackage{float}

\usepackage[
   backend=bibtex,
  citestyle=numeric-comp,
  style=chem-rsc,
    maxbibnames=1,
    sorting=none
]{biblatex}

\AtBeginBibliography{\footnotesize}
\usepackage{hyperref}
\hypersetup{
    colorlinks = true,
    allcolors=blue,
}
\usepackage{abstract}

\usepackage{titlesec}
\titlespacing\section{0pt}{12pt plus 4pt minus 2pt}{8pt plus 4pt minus 2pt}
\titleformat{\section}[block]{\scshape\filcenter}{}{1em}{}

\titlespacing\subsection{0pt}{12pt plus 4pt minus 2pt}{8pt plus 4pt minus 2pt}
\titleformat{\subsection}[block]{\scshape\filcenter}{\thesubsection}{1em}{}

\usepackage{upgreek}
\usepackage{mathptmx}

\newcommand\taba[1][0.14cm]{\hspace*{#1}}
\usepackage{tabularx}
\usepackage{changepage}
\usepackage[version=4]{mhchem}
\usepackage{gensymb}
\usepackage{textcomp}
\begin{document} 

\twocolumn[{
\begin{flushleft}
\Large\textbf{Low-damage electron beam lithography for nanostructures on \ce{Bi2Te3}-class topological insulator thin films}
\end{flushleft}

\begin{flushleft}
{Molly P. Andersen$^{1,2}$, Linsey K. Rodenbach$^{2,3}$, Ilan T. Rosen$^{2,4,a}$, Stanley C. Lin$^{5}$, Lei Pan$^{6}$, Peng Zhang$^{6}$, Lixuan Tai$^{6}$, Kang L. Wang$^{6}$, Marc A. Kastner$^{2,3,7}$, David Goldhaber-Gordon$^{2,3,b}$}
\end{flushleft}

\begin{flushleft}
\footnotesize{$^1$\textit{Department of Materials Science and Engineering, Stanford University, Stanford, California 94305, USA}}\\
\footnotesize{$^2$\textit{Stanford Institute for Materials and Energy Sciences, SLAC National Accelerator Laboratory, 2575 Sand Hill Road, Menlo \taba Park, California 94025, USA}}\\
\footnotesize{$^3$\textit{Department of Physics, Stanford University, Stanford, California 94305, USA}}\\
\footnotesize{$^4$\textit{Department of Applied Physics, Stanford University, 348 Via Pueblo Mall, Stanford, CA 94305, USA}}\\
\footnotesize{$^5$\textit{Stanford Nano Shared Facilities, Stanford University, Stanford, California 94305, USA}}\\
\footnotesize{$^6$\textit{Department of Electrical and Computer Engineering, Department of Physics and Astronomy, University of California, Los Angeles, California 90095, USA}}\\
\footnotesize{$^7$\textit{Department of Physics, Massachusetts Institute of Technology, 77 Massachusetts Avenue, Cambridge, MA 02139, USA}}\\
\footnotesize{$^a$\textit{Present address: Research Laboratory of Electronics, Massachusetts Institute of Technology}}\\
\footnotesize{$^b$To whom correspondence should be addressed; E-mail: \texttt{goldhaber-gordon@stanford.edu}}
\end{flushleft}


\begin{abstract}
Nanostructured topological insulators (TIs) have the potential to impact a wide array of condensed matter physics topics, ranging from Majorana physics to spintronics. However, the most common TI materials, the \ce{Bi2Se3} family, are easily damaged during nanofabrication of devices. In this paper, we show that electron beam lithography performed with a 30 or 50~kV accelerating voltage -- common for nanopatterning in academic facilities -- damages both nonmagnetic TIs and their magnetically-doped counterparts at unacceptable levels. We additionally demonstrate that electron beam lithography with a 10 kV accelerating voltage produces minimal damage detectable through low-temperature electronic transport. Although reduced accelerating voltages present challenges in creating fine features, we show that with careful choice of processing parameters, particularly the resist, 100~nm features are reliably achievable. 
\end{abstract}
}]

\section{Introduction}

The discovery of topological insulators (TIs) has introduced exciting new directions in condensed matter physics, including dissipationless edge conduction without an applied magnetic field~\cite{chang2013,okazaki2021}, new platforms for spintronics~\cite{jamali2015,han2017,wang2017}, and even the possibility of Majorana fermions, which have been proposed as a basis for qubits~\cite{fu2008,wang2015}. The tetradymites \ce{Bi2Se3}, \ce{Sb2Te3}, and \ce{Bi2Te3} were among the first topological insulators predicted theoretically~\cite{zhang2009}. Successful thin film growth of these materials and experimental confirmation of their hallmark spin-momentum-locked, massless Dirac dispersion soon followed~\cite{zhang2009,chen2009}. Tuning the Fermi level was achieved by alloying Bi and Sb on cation sites, and in some cases Te and Se on the anion sites, to form materials like \ce{(Bi,Sb)2Te3} (BST)~\cite{zhang2011} and \ce{(Bi,Sb)2(Te,Se)3} (BSTS)~\cite{taskin2011}. The ability to position the Fermi level at or near the Dirac point, where physics related to the topological surface states is most visible, along with demonstration of the quantum anomalous Hall effect in magnetically doped compounds of BST~\cite{chang2013,chang2015}, has made BST and BSTS some of the most widely studied TI materials in the field.

However, starting with carefully optimized as-grown films does not guarantee fabricated devices will retain the original film's electronic qualities. The BST family is known to be easily damaged during processing. For example, tellurium can evaporate from the film if processing temperatures are not kept low. Furthermore, tellurium will preferentially oxidize upon prolonged air exposure or oxygen plasma cleaning~\cite{bando2000,volykhov2016,volykhov2018}. Nonetheless, many interesting micron-scale devices have been produced using careful choice of processing parameters~\cite{jamali2015,mahoney2017,okazaki2021}.

Scaling device features to the nanoscale -- critical for exploration of Majorana physics, for example -- introduces new challenges.  To pattern features below a few microns, most academic fabrication facilities use electron beam lithography (EBL) with 25-100~kV accelerating voltages. It has been standard to use such tools for TI nanostructures~\cite{williams2012,cho2012,dipietro2013,sochnikov2013,galletti2014,kurter2015,stehno2016,charpentier2017,chen2018,kunakova2019,kunakova2020}, although some EBL-free alternative methods have been proposed~\cite{schuffelgen2017}. In this paper we demonstrate that 30 and 50~kV accelerating voltages cause unacceptable damage to TI thin films. We also show that EBL with lower-energy 10~kV electron beams does not significantly change the electronic properties of TI films. Unfortunately, lower-energy EBL makes it difficult to reliably produce features on the order of 100~nm because of strong small-angle scattering of electrons passing through the resist. We provide guidance for generating 100~nm features in a liftoff-based fabrication scheme. Combining our guidance with other recommendations to reduce fabrication-related damage in TI devices~\cite{breunig2022} can enable production of nanostructured devices with minimal electrical sample degradation~\cite{rosen2021,wu2021}. 

\section{Methods}

\subsection{Materials growth}

Three different materials were used in this paper. Film 1 was composed of 8~quintuple layer (QL) \ce{(Bi_{0.5}Sb_{0.5})2Te3}; both Film 2 and Film 3 were composed of 6 QL \ce{(Cr_{0.12}Bi_{0.26}Sb_{0.62})2Te3}. All three films were grown on epi-ready semi-insulating GaAs (111)B substrates in an ultra-high vacuum Perkin-Elmer molecular beam epitaxy (MBE) system. Before growth, the substrates were loaded into the MBE chamber and pre-annealed at a temperature of 670\degree~C in a Te-rich environment to remove the oxide on the surface. During growth of Film 1, high-purity Bi, Sb and Te were evaporated from standard Knudsen cells respectively. The substrate was kept at 215\degree~C. During growth of Films 2 and 3, high-purity Cr, Bi, Sb and Te were evaporated from standard Knudsen cells respectively. The substrate was kept at 200\degree~C.


\subsection{Hall bar fabrication and measurement}

\begin{figure*}[t!]
\centering
	\includegraphics[width=17cm]{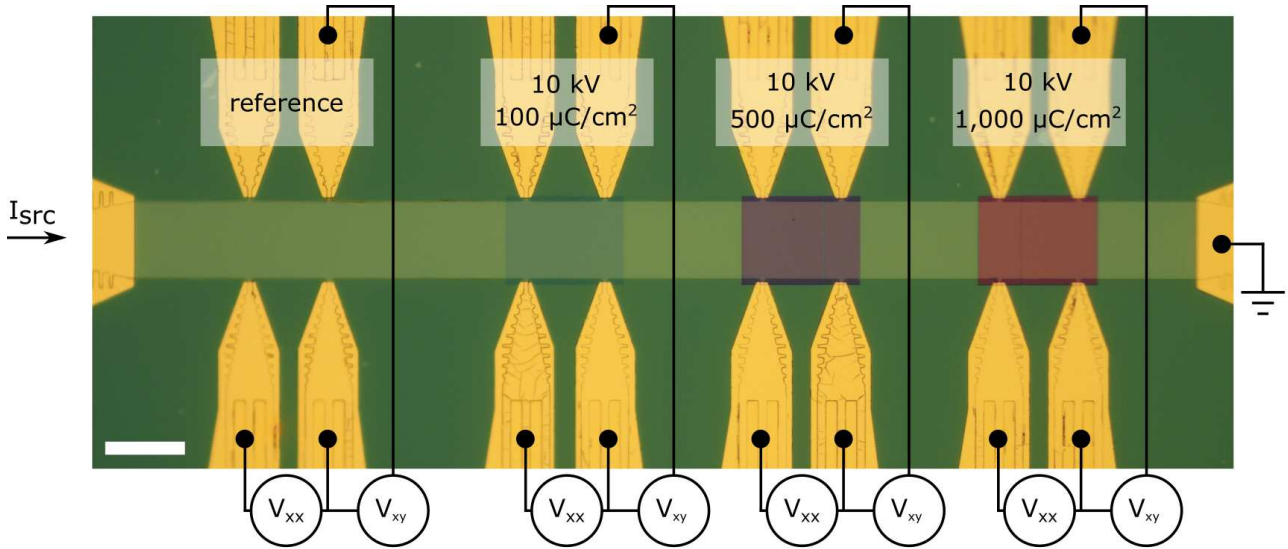}
	\caption{Optical image of a Hall bar device fabricated on Film 2, taken immediately after the device was exposed to electron beams with the undeveloped resist still in place. Contacts on the left and right side of the device were used to source ($I_{src}$) and drain current. Four quartets of contact pairs allow longitudinal ($V_{xx}$) and Hall ($V_{xy}$) voltage measurements of four isolated regions of the Cr-BST Hall bar. These four isolated regions were either left unexposed to electron beams (reference region) or exposed with a 10~kV electron beam at various doses, as indicated. Due to changes in the optical properties of PMMA after exposure, regions exposed to electron beams are visible as the colored rectangles over the device mesa. Scale bar: 80~$\mu$m.}
	\label{fig-device}
\end{figure*}

To study the effect of electron beam exposure on electronic transport in BST and chromium-doped, magnetically-ordered BST (Cr-BST), devices were fabricated on two chips of Film 1 (BST) and one chip of Film 2 (Cr-BST).  Devices consisted of long Hall bars with six, eight, or ten voltage contacts on each of the top and bottom edge. As shown in Figure~\ref{fig-device}, regions of the device mesa probed by two sets of contact pairs were locally exposed to electron beams; this enabled longitudinal and Hall transport measurements of isolated regions of each Hall bar. Between the two BST chips, three long Hall bars--each with three quartets of longitudinal and transverse contact pairs--were fabricated (Figure~\ref{sfig-TI-pics}).  The Cr-BST chip featured two long Hall bars, one with four and one with five quartets of contact pairs (Figure~\ref{sfig-QAH-pics}). On the BST chips, three regions of the BST material were left unexposed to electron beams; four regions were exposed with doses ranging 100--200~$\mu$C/cm$^2$ with a 10~kV accelerating voltage; two regions were exposed with 300 or 600~$\mu$C/cm$^2$ doses with a 30~kV accelerating voltage. On the Cr-BST chip, a total of two Cr-BST regions were left unexposed to electron beams; five regions were exposed at 10~kV with doses that ranged from 100--1,000~$\mu$C/cm$^2$; two regions were exposed at 50~kV with 500~$\mu$C/cm$^2$ and 2,500~$\mu$C/cm$^2$ doses. Higher exposure doses were selected at higher accelerating voltages to account for the larger clearing dose for resist at these voltages, due in turn to the lower scattering cross sections at higher accelerating voltages. Specific accelerating voltages were chosen based on the capabilities of the tools employed for patterning~\cite{sup}.

Devices on all chips were fabricated as follows: first, photolithography and argon ion milling were used to define device mesas on the as-grown BST or Cr-BST film. Next, Ti/Au contacts were added using photolithography, electron beam (e-beam) metal evaporation, and liftoff. Each chip was then again coated with polymer resist, in this case  poly(methyl methacrylate) (PMMA), commonly used for electron beam lithography. Isolated regions of the Hall bars were then exposed to electron beams to simulate the exposure that would occur during electron-beam lithography. Device micrographs taken at this point in the fabrication process are shown in the supplemental materials (Figures~\ref{sfig-TI-pics},~\ref{sfig-QAH-pics}). Electrostatic gates were then added to the Cr-BST devices as follows: After removing the PMMA with a solvent rinse and globally depositing a 1~nm aluminum seed layer with e-beam evaporation, 40~nm of alumina was deposited globally using atomic layer deposition to act as a gate dielectric. Photolithography, e-beam evaporation, and liftoff were used to define the metallic top gate. Finally, another photolithography step defined a mask through which alumina covering the contact pads was removed by a wet etch. Standard low-frequency electronic transport measurements of the BST (Cr-BST) Hall bars were performed at a base temperature of 1.55~K (30~mK). 

Throughout the fabrication process, care was taken to avoid thermal or chemical damage to the native Cr-BST and BST films. As noted in literature surrounding fabrication of HgTe devices~\cite{daumer2003,baenninger2012}, standard processing temperatures can impact the quality of fragile topological materials. Following past work with BST and Cr-BST~\cite{mahoney2017,fox2018,rosen2021,rodenbach2022,rosen2022}, all photo- and e-beam resist bakes were performed at 80\degree~C, not 120-180\degree~C as is typical for resist bakes on less fragile substrates. Full details of the fabrication procedure and electrical measurements, as well as discussion of processing differences between the BST and Cr-BST chips, can be found in the supplemental materials.

\subsection{Point spread function simulations}

Point spread functions (PSFs) describe the spatial pattern of energy density deposited into a particular resist/sample/substrate stack after exposure to an electron beam point source. PSFs shown in Figure~\ref{fig3}(a,b) and in the supplemental materials were simulated with the GenISys TRACER Monte Carlo simulation tool. The stack through which electron trajectories were simulated included, from the top down: (1) 150~nm PMMA (2) 8~nm \ce{Sb2Te3} (3) 0.5~mm GaAs. The electron injection energy was fixed at the accelerating voltage times one electron charge. For each accelerating voltage, one million trajectories were simulated. 

\subsection{Lithography tests}

Demonstrations of 100~nm features patterned by 10~kV EBL were performed on four chips cut from the Cr-BST Film 3 and on two chips of bare GaAs. Each chip was cleaned with acetone and isopropanol rinses. After cleaning, each chip was spun at 4000~rpm with a PMMA resist and baked at 80\degree~C. Resist specifications and bake times are listed in Table~\ref{table1}. After baking, all four chips were loaded together into a Raith VOYAGER electron beam lithography system. Exposures were performed with a 10~kV beam accelerating voltage and a 66.9~pA beam current. A 5~nm step size was used for area writes. For writes of single-pixel lines (SPLs), either 3, 5, or 8~nm step sizes were used~\cite{sup}. Patterns on the Cr-BST film targeted thin exposed lines as well as the inverse pattern of thin resist bridges - where two large areas are written as close as possible to one another while leaving some resist unexposed between them. On the GaAs chips, writes targeted the same goals, but were extended laterally over 1~mm to enable cleaving across the key features to image resist profiles.

After exposure, Cr-BST and GaAs chips were developed at ambient temperature with 55~s 1:3 methyl isobutyl ketone:isopropanol / 20~s isopropanol and immediately blown dry. Exposed regions of the Cr-BST chips were then metallized with 50~nm e-beam-evaporated Al after a 10~s \textit{in situ} Ar ion etch, followed by liftoff in acetone with sonication. The GaAs chips were instead sputter coated with 4.5~nm 60/40 Au/Pd and then cleaved across the features of interest~\cite{sup}. Scanning electron micrographs were then taken of all samples. The Cr-BST samples were imaged top-down to extract the horizontal length scales of the critical features; GaAs samples were imaged at an angle to view cross-sectional resist profiles. 

\section{Results and Discussion}

\subsection{Electron beam-induced damage to topological insulator films}

Electron beam exposures are described by an accelerating voltage, used to accelerate electrons towards the sample, and a charge dose, describing the electron fluence that crosses the surface of the sample's resist. The clearing dose describes the electron fluence necessary to fully expose a given resist stack, meaning that the exposed region of resist will be fully dissolved during development, and is typically the minimum dose needed for a successful EBL exposure. The magnitude of the clearing dose is determined by the resist thickness, how strongly the electron beam interacts with the resist, and any back-scattering off of the sample, which effectively increases the electron flux through the resist. Clearing dose can also be affected by choice of development conditions. Throughout this work, we use the developer 1:3~methyl isobutyl ketone:isopropanol/isopropanol. The clearing dose tends to increase roughly linearly with accelerating voltage since scattering cross-sections decrease as the electron beam energy increases. 

Although electron beams can be extremely narrow ($\sim8$~nm) as they enter a sample's resist coating, they scatter and spread out as they pass through the resist and into the sample below. This broadening causes proximity effects, by which the effective dose is higher towards the center of large area writes due to overlap between adjacent exposures. In contrast, near corners and edges, as well as in very thin or single-pixel line writes, the effective dose is reduced since there are fewer adjacent points exposed. As a result of these proximity effects, the doses required for thin lines or near corners and edges can be much higher than clearing doses in the center of large area writes. 

\begin{figure}[h t!]
\centering
	\includegraphics[width=5cm]{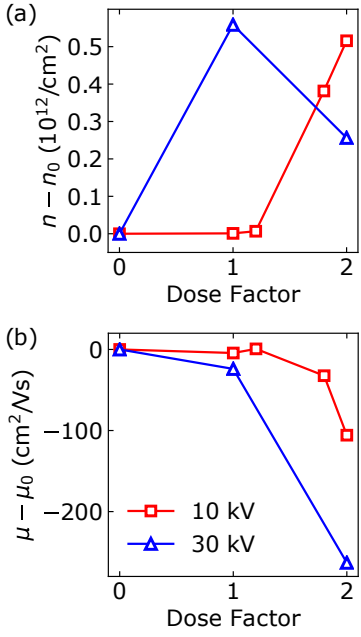}
	\caption{Effects of electron beam exposure on nonmagnetic TIs. Changes in (a)~density and (b) mobility relative to a reference region unexposed to electron beams. Data are shown as a function of the dose factor, the ratio of the exposure dose to the clearing dose. For~10 kV exposures (red squares) the clearing dose is 100~$\mu$C/cm$^2$, and exposures range from 0-200~$\mu$C/cm$^2$. For 30~kV exposures (blue triangles) the clearing dose is 300 $\mu$C/cm$^2$, and exposures range from 0-600~$\mu$C/cm$^2$. Raw data are presented in Figure~\ref{sfig-TI-damage}.}
	\label{fig1}
\end{figure}

As discussed above, to study the effects of electron beam exposure on nonmagnetic TIs, several Hall bars were fabricated on a BST film (Film 1). Density $n$ and mobility $\mu$ of each locally exposed region were extracted from Hall and longitudinal resistance measurements at $T\approx1.5$~K. As shown in Figure~\ref{fig1}(a,b), for exposure at either 10~kV or 30~kV the density increased (indicating electron doping) and the mobility decreased (indicating increased disorder) as the dose of electrons was increased. However, exposures at 10~kV caused substantial deviations from the reference unexposed regions only at doses higher than the clearing dose. In contrast, regions exposed at 30~kV exhibited substantial density increases and mobility decreases even at the clearing dose. These data indicate that clearing dose exposures at 10~kV only minimally perturb electronic transport behavior in BST, whereas exposures at 30~kV significantly degrade the material. 

As discussed above, Hall bars with local electron beam exposure were also fabricated on a Cr-BST chip (Film 2). Magnetically-doped TIs, including Cr-BST, are known to host the quantum anomalous Hall effect (QAHE), a zero-magnetic-field analog of the quantum Hall effect~\cite{chang2013,chang2015}. Ferromagnetic order of the dopants breaks time reversal symmetry and opens a gap in the Dirac surface state of the TI. When the Fermi level is tuned to lie within this mass gap, the QAHE is observed. In electrical transport measurements the QAHE is seen as a vanishing longitudinal conductivity, $\sigma_{xx} = 0$, and a quantized Hall conductivity,  $\sigma_{yx}=\pm e^2/h$, whose sign is dependent on the Chern number $C=\pm1$ and tuned by the out-of-plane orientation of the sample's magnetization. A quantized value of the conductivity tensor ($\sigma_{yx}=\pm e^2/h$, $\sigma_{xx}=0$) corresponds to complete magnetization of the sample in the positive or negative out-of-plane direction and the corresponding Chern number $C=\pm1$. 

\begin{figure*}[t!]
\centering
	\includegraphics[width=17cm]{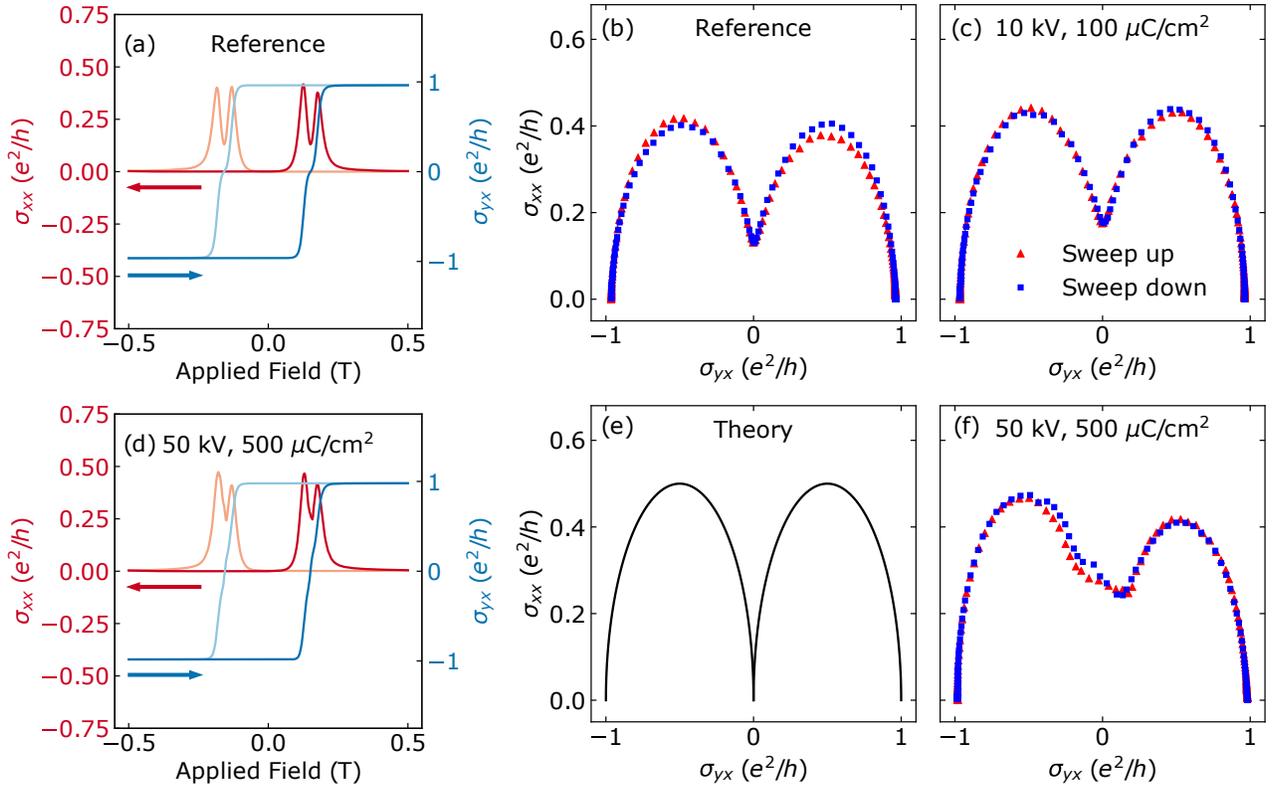}
	\caption{Effects of electron beam exposure on electronic transport in Cr-BST. Longitudinal conductivity $\sigma_{xx}$ (red, left axis) and Hall conductivity $\sigma_{yx}$ (blue, right axis) are shown as a function of applied magnetic field for (a) a reference region, unexposed to an electron beam and (d) a region exposed at the clearing dose with a 50~kV accelerating voltage. Magnetic field was swept both up (darker traces) and down (lighter traces) to generate hysteresis loops. The same longitudinal conductivity data are plotted parametrically as a function of Hall conductivity for (b) the reference region and (f) the region exposed at the clearing dose at 50~kV. Additional parametric plots are shown for (c) the same measurement performed on a region exposed at the clearing dose with a 10~kV accelerating voltage and (e)  theoretical behavior when the system is tuned explicitly through a trivial insulating phase upon magnetization reversal. In all parametric plots, data acquired while sweeping field up (down) are shown as red triangles (blue squares).}
	\label{fig2}
\end{figure*}

After magnetizing the Cr-BST sample in a 0.5~T magnetic field normal to the plane of the device and tuning the Fermi level to the center of the magnetic exchange gap via electrostatic gating, the QAHE was observed for all devices, independent of electron beam exposure~\cite{sup}. Figure~\ref{fig2}(a) shows a typical QAHE hysteresis loop acquired for a region of a Hall bar that was not exposed to electron beams as the magnetic field normal to the plane of the device is swept back and forth between $-0.5$~T and $0.5$~T, well past the film's coercive field of $0.2$~T. The left (red) axis plots the longitudinal conductivity $\sigma_{xx}=\rho_{xx}/(\rho_{xx}^{2}+\rho_{yx}^{2})$ which approaches 0~$e^2/h$ away from the coercive field. The right (blue) axis plots the Hall conductivity $\sigma_{yx}=\rho_{yx}/(\rho_{xx}^{2}+\rho_{yx}^{2})$, which switches between $\sigma_{yx}=\pm e^2/h$ at the coercive field. Conductivities ($|\sigma_{yx}|< e^2/h$, $\sigma_{xx}>0$) occur when the sample magnetization is not homogeneous and out-of-plane, and are observed as the magnetization reverses at external magnetic fields close to the coercive field $\pm 0.2$~T. Figure~\ref{fig2}(b) replots the data from (a) in a standard parametric plot of $\sigma_{xx}$ against $\sigma_{yx}$, which visualizes the flow of the QAHE across the $C=\pm 1$ topological phase transition at magnetization reversal~\cite{fox2018,pan2020}. 

When the magnetization is reversed in Cr-BST, one of two patterns is observed in conductivity measurements: (1) explicit tuning through a trivial insulating phase $C=0$, which is characterized by $\sigma_{xx}\sim0$, ${\sigma_{yx}}\sim0$  (Figure~\ref{fig2}(b,e)), and (2)~directly tuning between $C=\pm 1$ without observable trivial insulating behavior~\cite{pan2020,wang2014,feng2015}. Which of these two behaviors occurs is thought to correlate with the thickness of the Cr-BST film, possibly as a result of competition between the topologically nontrivial magnetic exchange gap and a topologically trivial gap formed by hybridization of surface states in films less than $\sim$6~nm thick~\cite{pan2020}. Fabrication techniques that alter a sample's behavior at the topological phase transition risk convoluting transport features of the physics of interest with damage-related phenomena or destroying them entirely. For example, proposals to study chiral Majorana edge modes at interfaces between quantum anomalous Hall and superconducting materials rely on careful control over a  $C=\pm 1 \rightarrow C=0$ phase transition~\cite{chung2011,wang2015}. 

The behavior across the topological phase transition at this magnetization reversal was observed to evolve with both dose and accelerating voltage. As shown in the parametric plots of Figure~\ref{fig2}, regions exposed with a 10~kV accelerating voltage maintain the same qualitative behavior as the reference region across magnetization reversal. However, regions exposed at 50~kV exhibit an asymmetric transition between magnetization orientations. This distinction between how the reference region and regions exposed at 50~kV pass through the topological phase transition at the coercive field can also be seen in the hysteresis loop of Figure~\ref{fig2}(d), where the asymmetry in $\sigma_{xx}$ across the coercive fields is responsible for the asymmetry in Figure~\ref{fig2}(f). The behavior of the regions exposed at 50~kV represents a substantial deviation from the as-grown film behavior and is clear evidence of electron beam-induced damage. Data from additional regions of the Cr-BST Hall bars are presented in the supplemental materials.

Additionally, in areas exposed with a 10~kV accelerating voltage, the optimal gate voltage shifts nearly linearly with dose~\cite{sup}. Since the optimal gate voltage reflects the doping and number of mid-gap states, this shift indicates that electron doping and/or population of defect states scales with the electron dose, as expected. Areas exposed at 50~kV did not display a monotonic relationship between dose and doping~\cite{sup}. We speculate that this difference indicates that damage mechanisms are different at the different accelerating voltages, but further study is needed to clarify the relationship between dose and damage.

\subsection{Optimizing 10~kV electron-beam lithography for 100~nm features}

We have shown that damage to the \ce{Bi2Te3} family of materials during EBL significantly alters their electronic properties, but that this can be mostly avoided by using a 10~kV accelerating voltage. Yet patterning nanostructures at such a low accelerating voltage presents challenges. At standard accelerating voltages (25-100~kV), electrons undergo minimal small-angle forward scattering as they pass through the resist layer~\cite{sup} and tend to pass straight through the resist~\cite{takigawa1983,zhou2006,chen2015}. In contrast, at 10~kV electrons undergo substantial small-angle forward scattering as they initially pass through the resist layer. As a result, the electron beam broadens considerably as it travels through the resist~\cite{takigawa1983,takigawa1991,zhou2006,olkhovets1999}.

\begin{figure*}[t!]
\centering
	\includegraphics[width=17cm]{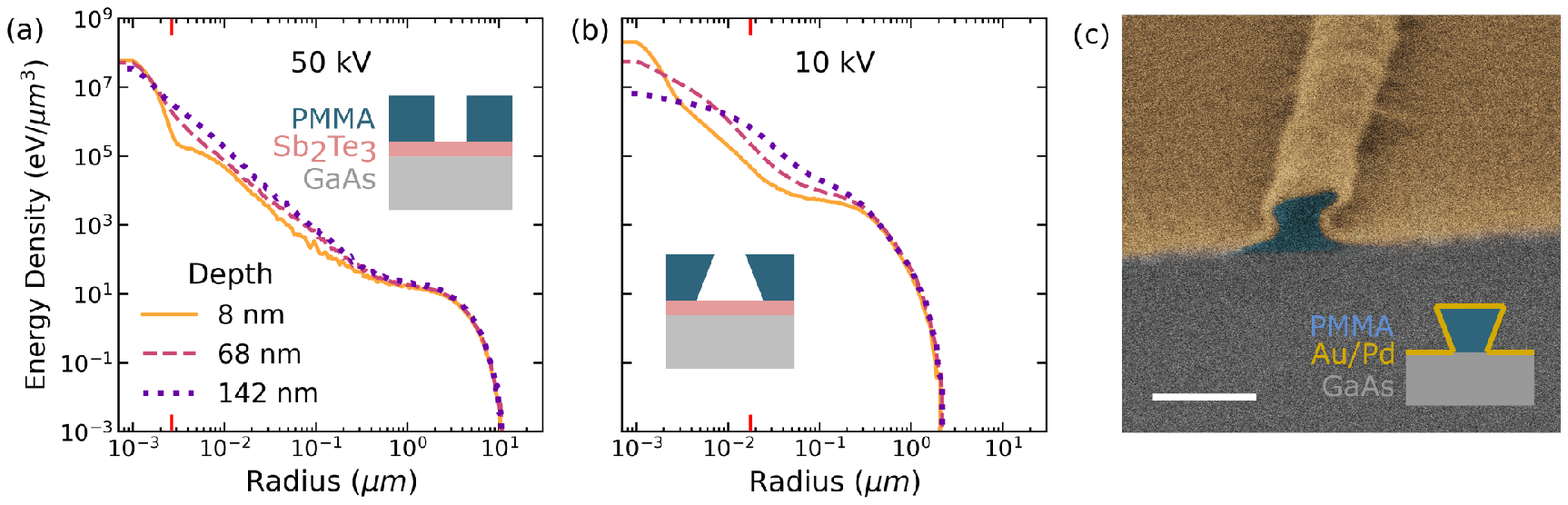}
	\caption{(a,b) Simulated volumetric energy density deposited into a 150~nm layer of PMMA / 8~nm \ce{Sb2Te3} / 500~$\mu$m GaAs stack as a function of distance from the center of the electron beam. Data with an accelerating voltage of 50~kV [10~kV] are shown in (a)[(b)]. Line cuts are shown at three different depths from the top of the PMMA layers: 8~nm from the top (yellow line), 68~nm from the top (pink dashed line), and 142~nm (8~nm) from the top (bottom) of the resist layer (purple dotted line). Red markers on the axes indicate the radius at which deposited energy density decreases by an order of magnitude for the 142~nm traces, closest to the sample's surface. Insets: expected cross-sectional profiles after exposure and development. Additional PSFs and discussion are presented in the supplemental materials. (c)~False-color scanning electron micrograph of a cross-sectional profile of a thin resist bridge (center) after exposing the surrounding regions with a 10~kV electron beam and developing. Blue: PMMA. Gold: Au/Pd charge-neutralizing layer to aid in imaging. Grey: GaAs substrate. Scale bar: 100~nm.}
	\label{fig3}
\end{figure*}

Figure~\ref{fig3}(a,b) shows simulations of energy density deposited into resist as a function of distance from the center of the electron beam at three different depths within a 150~nm PMMA resist layer. As shown in Figure~\ref{fig3}(a), with a 50~kV accelerating voltage most of the energy is deposited within 10~nm of the center of the beam throughout the resist. With a 10~kV accelerating voltage, the energy is again deposited within about 10~nm laterally of the beam center at the top of the resist layer but within a much broader 100~nm lateral distance from the beam center near the bottom of the resist. This difference in beam broadening is marked by the red marker on the x-axes of Figures~\ref{fig3}(a,b), which indicate the radius at which the energy density has dropped by a factor of ten relative to the energy density at the center of the beam at the surface of the \ce{Sb2Te3} (depth 142~nm). Insets in Figure~\ref{fig3}(a,b) sketch resultant cross-sectional resist profiles after development. At 50~kV accelerating voltage, the deposited energy remains close to the center of the beam, therefore resist sidewalls remain vertical throughout the resist layer. At 10~kV, however, small-angle scattering causes the deposited energy to broaden as the beam passes through the resist, so the sidewalls taper away from the beam center. Figure~\ref{fig3}(c) shows a scanning electron micrograph of a resist profile after exposing resist at 10~kV accelerating voltage, developing, and then depositing a thin metal layer to enhance image contrast; inwardly-slanted sidewalls are clearly visible. 

Slanted sidewalls make $=<$100 nm features challenging to produce with a 10~kV accelerating voltage. Narrow resist bridges can completely pinch off and delaminate close to the substrate~\cite{sup}. Even if this does not occur, dramatic undercuts can reduce the mechanical stability of thin bridges. These issues limit resist choices to thinner layers. Unfortunately, thin resist limits subsequent fabrication steps: liftoff and dry etches fail if the resist layer is too thin. 

Nevertheless, 100~nm features are consistently achievable with a 10~kV accelerating voltage through careful processing and patterning choices, even when bake temperatures must be kept low to avoid damage to sensitive materials. To demonstrate some important considerations, we fabricated sub-100~nm gap and line features using EBL followed by liftoff of 50~nm aluminum, varying several resist-related parameters between samples. Table~\ref{table1} shows PMMA resist choices and bake times for four chips C1-C4. The resist labels of 495 or 950 describe the PMMA molecular weight in units of $10^3$~g/mol. All resists chosen here are diluted in anisole; the notation AX in Table~\ref{table1} indicates an X\% anisole dilution.  All bakes were performed in ambient atmosphere at a reduced temperature of 80\degree~C to avoid thermal degradation of samples. The thinnest resist, used on C1 and C2, is approximately 130 nm thick, roughly as thin as possible for liftoff of 50~nm metal. 

\begin{table}[ht!]
 \caption{\label{table1} Resist and bake time choices for four chips of Film~3 used to test EBL resolution with writes at 10~kV. All other processing parameters are described in the Methods section; notably, resist bake temperatures were limited to 80\degree~C.} 

\begin{tabular}{@{}ccc}

Sample & Resist (PMMA)& Bake Time (min.)\\

C1 & 950 A3 & 5\\
C2 & 950 A3 & 30\\
C3 & 950 A5 & 5\\
C4 & 495 A4/950 A3 bilayer & 5/5\\

\end{tabular}

\end{table}

As discussed above, gaps were patterned by exposing two large areas separated by a nanometer-scale nominal gap size at varied doses. After development, a thin resist bridge (similar to Figure~\ref{fig3}(c)) remained. Metal liftoff inverts the pattern, leaving behind two large metallized regions separated by a gap. This has long been a typical fabrication flow for liftoff-based weak link Josephson junctions~\cite{williams2012,veldhorst2012,veldhorst2012_apl}. Figure~\ref{fig4}(a) shows representative results for gaps fabricated on C1-C4; full results are shown in the supplemental materials.

Measured gap sizes down to 80~nm at best and 100~nm on average were attained. Chip 2, which featured a longer bake and thin resist, produced the best results overall and most consistently across different doses. C1, which featured a short bake and thin resist, also produced consistently good results, though slightly worse than C2. C3 and C4, with thicker resist stacks, produced worse results than C1 or C2, but still reached gaps less than 100~nm in some cases. Nominal gap sizes -- the gap sizes of the pattern -- were larger than measured gap sizes by roughly a factor of two, but the ratio varied between chips, doses, and the nominal gap sizes. 

Thin lines were patterned by exposing a single pixel line (SPL) with the electron beam at various doses as discussed above. After development, metal deposition, and liftoff, a thin line of metal remains (Figure~\ref{fig4}(b) inset). Such patterns are important for line gates or thin etched trenches~\cite{cho2012,dipietro2013}. Figure~\ref{fig4}(b) shows results for C1 and C2 with a 3~nm electron beam step size. Measured linewidth roughly increases with dose, as expected; past the clearing dose, features tend to broaden with increasing dose. For both C1 and C2, lines down to about 100~nm wide were attained at the lowest successful line dose (C1: 1,300~pC/cm; C2: 1,200~pC/cm). Additionally, doses up to 50\% larger produced consistent 100-120~nm lines. Additional results for a 5 and 8~nm step size are shown in Figure~\ref{sfig-TI-lines}. Among the doses tested here, no viable lines were produced in Chips C3 and C4 because the thicker resists used on these chips require higher doses to fully expose deeper regions of the resist due to small-angle forward scattering. 

The best results were obtained when using the thinnest resist layers, with 80-120~nm gaps and lines readily obtained within specific windows of dose and patterning parameters (nominal gap size, electron beam step size).

Thicker resist stacks produced some comparable gap sizes, but for a narrower range of processing parameters. The inferior performance of thicker resist at 10~kV accelerating voltage is consistent with the expectation of less mechanically-stable sidewalls and concomitantly increased sensitivity to patterning parameters. Comparing C1 and C2, increased bake times improved results. We speculate that longer bake times partially compensated for the low bake temperature used of 80\degree~C, well below the $\sim$105\degree~C glass transition of PMMA.
 
\begin{figure}[t!]
\centering
	\includegraphics[width=6cm]{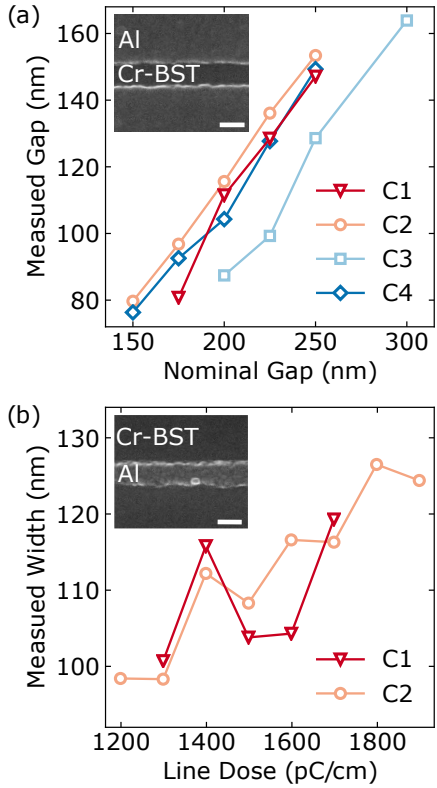}
	\caption{Feature sizes for gaps and lines written with a 10~kV accelerating voltage. Insets: SEM images of smallest feature shown for C2; scale bar: 100 nm. (a) Measured gap width as a function of nominal gap width for C1-C4. Data were acquired for a range of doses; data shown here correspond to the dose at which the narrowest gaps were observed for each chip (C1: 110~$\mu$C/cm$^2$. C2: 100~$\mu$C/cm$^2$. C3: 130~$\mu$C/cm$^2$. C4: 120~$\mu$C/cm$^2$.). The full data set is presented in the supplemental materials. (b)~Measured line widths as a function of line dose for chips C1 and C2 for single pixel lines written with a 3~nm step size. Data for 5 and 8~nm step sizes are shown in the supplemental materials.}
	\label{fig4}
\end{figure}

\section{Conclusions}

In this paper, we demonstrated that electron beam lithography damages the topological insulator BST and its magnetic analogue Cr-BST. In the nonmagnetic material, we showed that exposure to electron beams increased electron density and decreased mobility. In Cr-BST, we found that electron beam exposure altered the electronic transport of the material at the topological phase transition associated with magnetization reversal. Whereas here we measured degradation of bulk transport properties in regions uniformly exposed to electron beams, in real nanopatterned devices the electron dose received by the film varies spatially. In such devices, it becomes prohibitively challenging to separate fabrication-related materials damage from physics of interest. 

Although we found that the severity of the changes to the materials' electronic properties after exposure at 30 and 50~kV accelerating voltages (at doses appropriate for lithography) are unacceptable for many experiments on TI nanostructures, we showed that lithography at 10~kV imparts minimal change to bulk electronic properties of canonical topolocical materials. Further study is required to elucidate the accelerating-voltage-dependent microscopic mechanisms of electron beam-induced damage, which may include creation of atomic point defects, or other structural materials changes. 

Although increased small-angle scattering makes fine features difficult to pattern at 10~kV compared to at higher voltages, we demonstrated a window of processing parameters that generate 100~nm-wide lines or gaps after metal liftoff. Since 100~nm features are sufficient for many nanostructures of interest, we suggest that electron beam lithography at $\geq 30$~kV should be avoided in TI device fabrication. We further suggest that, when sub-100~nm features are required, electron beam lithography at intermediate accelerating voltages 15-25~kV could provide modest improvements in patterning resolution at the expense of small, but tolerable, increases in damage. Further work is required to quantify this trade-off. Additionally, alternative nanopatterning techniques have been developed, including selective area growth~\cite{kampmeier2016} and  stencil lithography~\cite{schuffelgen2017}, although these techniques require specialized equipment, and materials damage (from, for example, unwanted interface chemistry) has not been explicitly studied. 

\section*{Acknowledgements}

The authors thank R. C. Tiberio and E. J. Fox for useful discussions. M. P. A., L. K. R., I. T. R., M. A. K., and D. G.-G. were supported by the U.S. Department of Energy, Office of Science, Basic Energy Sciences, Materials Sciences and Engineering Division, under Contract DE-AC02-76SF00515. K. L. W. and L. T. were, in part, supported by the Army Research Office MURI program under Grant No. W911NF-16-1-0472 and by the National Science Foundation under Grant No. DMR1936383. Infrastructure and cryostat support were funded in part by the Gordon and Betty Moore Foundation through Grant No. GBMF3429. We thank NF Corporation for providing low-noise, high-input-impedance voltage preamplifiers. We acknowledge measurement assistance from colleagues at the National Institute of Standards and Technology. Part of this work was performed at the nano@Stanford labs, supported by the National Science Foundation under award ECCS-2026822. 

\section*{Data Availability Statement}

Point spread function simulations and transport data that support the findings of this study are available at http://doi.org/10.5281/zenodo.7549123. Scanning electron micrographs used to generate the data of Figure~\ref{fig4} are available from the corresponding author upon reasonable request. 

\section*{Author Declarations}

The authors have no conflicts to disclose.

\section*{Author Contributions}

\textit{Resources:} L. P., P. Z., L. T., and K. L. W.; \textit{Conceptualization:} M. P. A. and L. K. R.; \textit{Methodology:} M. P. A., L. K. R., I. T. R., S. C. L., L. P., P. Z., L. T., and K. L. W.; \textit{Investigation:} M. P. A. and L. K. R.; \textit{Formal Analysis:} M. P. A. and L. K. R.; \textit{Supervision:} M. A. K. and D. G.-G.; \textit{Writing:} M. P. A. with input from all authors.

\setlength\bibitemsep{0pt}
\printbibliography

\onecolumn
\begin{center}
\large\textbf{Supplementary Materials}
\end{center}
\setcounter{figure}{0}
\setcounter{section}{0}
\renewcommand{\thefigure}{S\arabic{figure}}
\renewcommand{\theequation}{S\arabic{equation}}

\section{Details of experimental methodology}

\subsection{BST Hall bar fabrication and measurements}

\begin{figure*}[ht!]
\centering
	\includegraphics[width=0.7\textwidth]{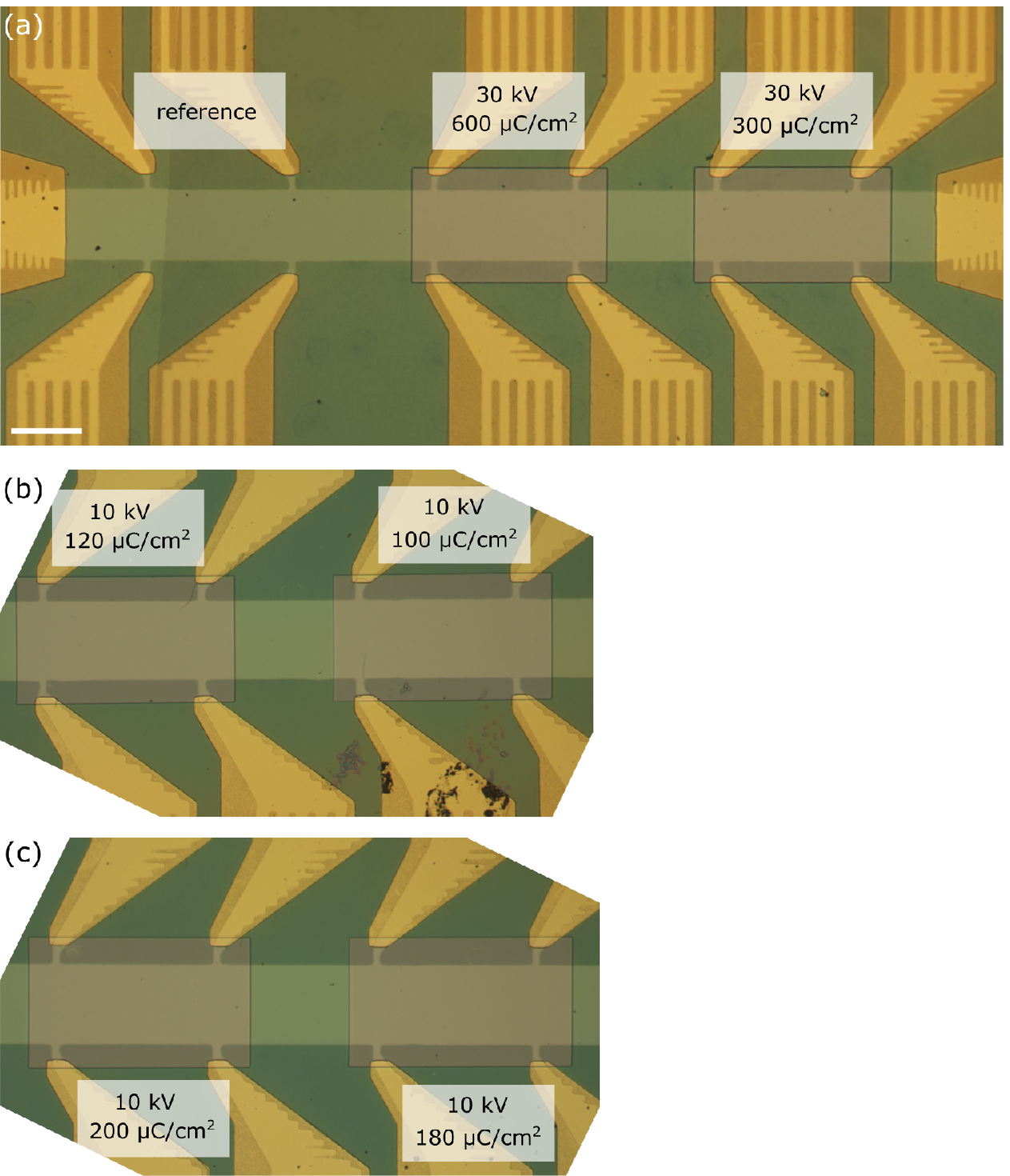}
	\caption{Microscope images of Hall bar devices fabricated on BST and used to produce the data in Figure~\ref{fig1}. Images were taken after electron beam exposure and development. (a) Long Hall bar with three quartets of contact pairs, with some regions (rectangles between middle and right contact pairs) exposed by a 30 kV electron beam and the indicated doses. Picture shown was stitched together from two separate images (break between images falls in the reference region). Scale bar: 80 $\mu$m. (b,c) Regions of two separate Hall bars exposed to a 10 kV electron beam at the indicated doses. Only the exposed regions are shown, but the full device geometries are identical to that shown in (a). Both images are irregular shapes because the original images were taken at an angle, and then cropped to show the regions of interest.}
	\label{sfig-TI-pics}
\end{figure*}

Two separate chips were cut from a single wafer of Film 1. On one chip,a Hall bar with three quartets of longitudinal and transverse contact pairs was fabricated as shown in Figure~\ref{sfig-TI-pics}(a)  to test the effects of 30 kV e-beam lithography. On the second chip, two Hall bars with structures identical to that shown in Figure~\ref{sfig-TI-pics}(a) were fabricated and exposed to 10 kV electron beams (exposed regions with various fluences are shown in Figure~\ref{sfig-TI-pics}(b,c)). 

To define the sample mesa, Hall bars were patterned with photolithography (SPR 3612 resist, 2 min 80\degree~C resist bake, phosphate salt developer followed by two water rinses) and etched in an Ar ion mill with an accelerating voltage of 300 V. Samples were cleaned with a 30 s acetone sonication and solvent rinse. Contacts were added with photolithography and electron beam (e-beam) evaporation of 5/80 nm Ti/Au after a 20 s in situ Ar pre-etch. Metal liftoff was performed with 20 s sonication in acetone and solvent rinse. The BST Hall bars were spin-coated with polymethyl methacrylate (PMMA) A5 950. Both chips were baked for 6 min at 80\degree~C. At this point, isolated regions of the Hall bars were exposed to electron beams. All electron beam exposures on BST devices were performed on a FEI Nova NanoSEM. Writes at 10 kV used a 220 pA beam and 10 nm step size. Writes at 30 kV used a 600 pA beam and 10 nm step size. After exposure, the samples were developed for 50 s in a 1:3 solution of ambient-temperature methyl isobutyl kethone:isopropanol and photographed (Figure~\ref{sfig-TI-pics}). The resist was then globally removed with a solvent rinse.

Electrical measurements of BST Hall bars were made in a variable temperature insert $^4$He refrigerator with a base temperature of $\sim$1.55 K. A Stanford Research Systems SR830 lock-in amplifier (SR830) was used to source 5 V across a 1 G$\Omega$ resistor to provide a 5 nA RMS current bias at $\sim$17 Hz . Measurements of longitudinal and Hall voltages as well as current were made by additional SR830s. Prior to measurement, voltages were amplified with using LI-75A voltage preamplifiers with a gain $10^2$. Density and mobility values were extracted from Hall measurements as a function of applied out-of-plane magnetic field ranging from  -0.5 T to +0.5 T. The Hall bar used for 30 kV accelerating voltage exposure was measured during a single cool-down. The two Hall bars used for 10 kV accelerating voltage exposure were measured during separate cool-downs.

\subsection{Cr-BST Hall bar fabrication and measurements}

\begin{figure}[ht!]
\centering
	\includegraphics[width=0.7\textwidth]{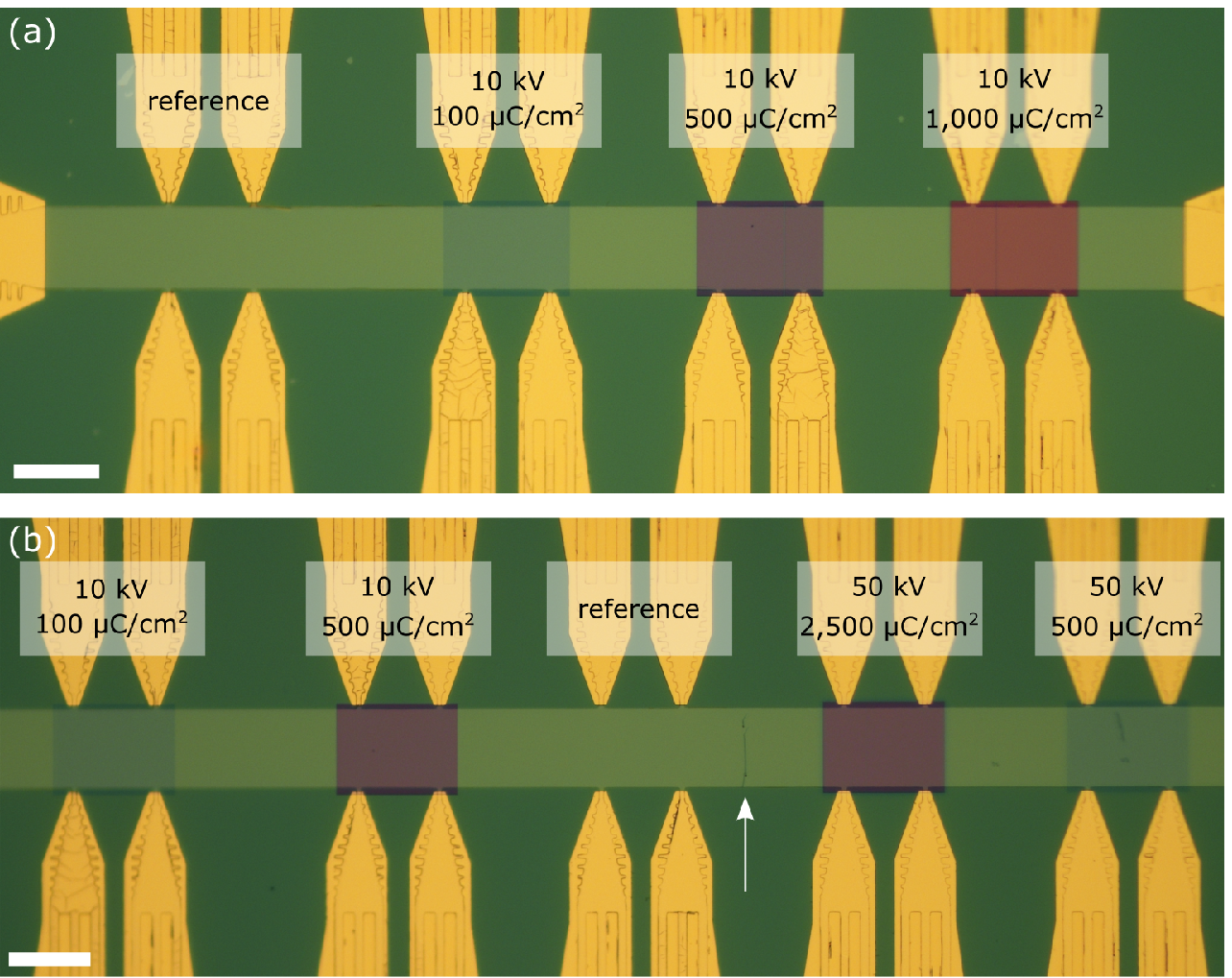}
	\caption{Microscope images of Hall bar devices fabricated on Cr-BST and measured to produce Figure~\ref{fig2}. Devices are shown after electron beam exposure, with undeveloped resist still in place. Exposed resist is visible as the colored squares over individual sets of contacts. These images are not false-colored; the exposed resist has colors different from unexposed resist. Exposure conditions are indicated for each exposed region. Scale bars, 80 $\mu$m. (a) Hall bar exposed exclusively at 10 kV and used for measurements in Figure~\ref{fig2}(a-c). (b) Hall bar exposed at both 10 and 50 kV and used for measurements in Figure~\ref{fig2}(d,f). The Hall bar mesa was damaged between the reference and 50 kV regions (indicated by arrow), but this damage had no noticeable impact on measurements in the QAHE regime.}
	\label{sfig-QAH-pics}
\end{figure}

Hall bar mesas were patterned on the bare Cr-BST Film 2 with photolithography (SPR 3612 resist, 5 min 80\degree~C resist bake, phosphate salt developer followed by two water rinses) and etched in an Ar ion mill with an accelerating voltage of 400 V. Samples were cleaned with a 20 s acetone sonication and solvent rinse. Contacts were added with photolithography and e-beam evaporation of 5/90 nm Ti/Au at a rate of 1 \AA/s after a 10 s in situ Ar pre-etch. Metal liftoff was performed with a 20 s sonication in acetone and solvent rinse. The Cr-BST chip was spin-coated with polymethyl methacrylate (PMMA) A5 950 and baked for 5 min at 80\degree~C.  At this point, isolated regions of the Hall bars were exposed to electron beams. All electron beam exposures on Cr-BST devices were performed on a Raith VOYAGER electron beam lithography system. Writes at 10 kV used a 60 pA current and 5 nm beam step size. Writes at 50 kV used a 5 nA current and 10 nm beam step size. After exposure, devices were photographed (Figure~\ref{sfig-QAH-pics}) but not developed. Instead, the resist was globally removed with a solvent rinse. A seed layer of 1 nm aluminum was globally deposited with e-beam evaporation at 0.3 \AA/s and allowed to oxidize in atmosphere. An alumina gate dielectric was then globally deposited with low-temperature (60\degree~C) atomic layer deposition. Top gate electrodes were defined with photolithography, and e-beam evaporation was used to deposit 5/95 nm Ti/Au top gate electrode metals at a rate of 1 \AA/s after 10 s \textit{in situ} Ar pre-etch. Liftoff was performed with 20 s acetone sonication and solvent rinse. Excess dielectric was removed over the Hall bar mesa contacts with a photolithographically-masked tetramethylammonium hydroxide-based wet etch (120 s/20 s/20 s Microposit CD-26 Developer/water/water). The resist was then globally removed with a solvent rinse. 

Electrical measurements of Cr-BST Hall bars were made in a dilution refrigerator with a base temperature of $\sim$30 mK after magnetizing the sample with a 0.5 T out-of-plane applied field. The measurement lines include low-pass RF filters as well as discrete RC filters at the mixing chamber stage. A Stanford Research 830 lock-in amplifier (SR830) was used to source 5 V across a 1 G$\Omega$ resistor to provide a 5 nA current bias at $\sim$5 Hz. Measurements of longitudinal and Hall voltages as well as current were measured by additional SR830s as well as Stanford Research 860 lockin amplifiers. Prior to measurement, voltages were amplified with either separate NF LI-75A voltage preamplifiers or a single NF multi-channel preamplifier, with a gain of $10^2$ V/V in either case. Current was amplified with an Ithaco 1211 current preamplifier with a gain of $-10^6$ V/A. Gate voltage was controlled with a Keithley Model 2400 Source-Measure Unit. Temperature was modulated with a heater on the mixing chamber stage of the dilution refrigerator, and was measured by a thermometer also on the mixing chamber stage. Each Hall bar on the Cr-BST chip was measured on a separate cool-down.

\section{Additional BST damage data}
\begin{figure}[H]
\centering
	\includegraphics[width=0.9\textwidth]{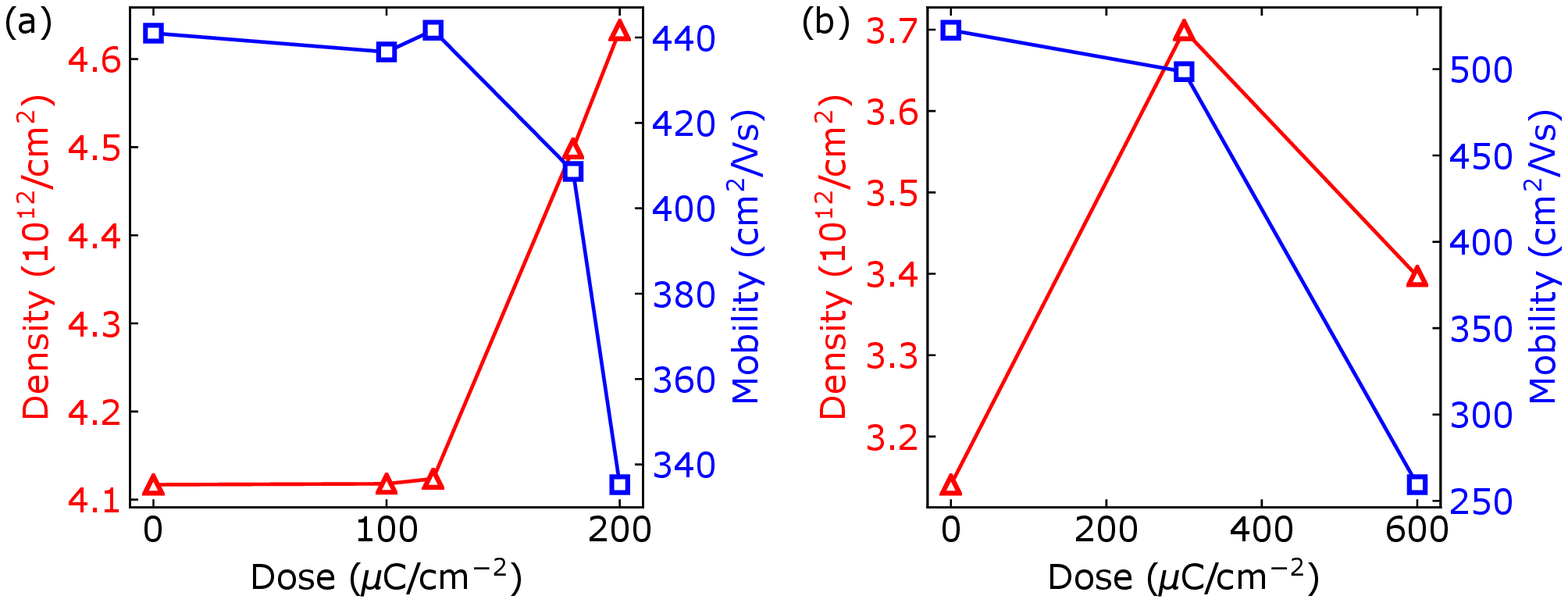}
	\caption{Raw data used to generate Figure~\ref{fig1}. Density (red triangles, left axis) and mobility (blue squares, right axis) as a function of dose for regions of a BST Hall bar exposed at (a) 10 kV or (b) 50 kV.}
	\label{sfig-TI-damage}
\end{figure}

\section{Additional Cr-BST damage data}

A temperature scale $T_0$ corresponding to an effective size of the magnetic exchange gap can be extracted from longitudinal and Hall transport measurements at elevated temperatures. As temperature is increased, some fraction of charge carriers are thermally excited across the magnetic exchange gap into surface state bands; these thermally activated carriers introduce dissipation into an otherwise dissipation-free system. Arrhenius fits of longitudinal conductivity can provide $T_0$ as the barrier to thermally activated dissipative conduction. At lowest temperatures, measurements typically plateau and diverge from a simple Arrhenius model due to deviation of electron temperature from lattice temperature and leakage currents in measurement electronics; as a result, an Arrhenius plus offset model ($\sigma_{xx} = a + b\exp{-T_0/T}$ with free parameters $a$ and $b$) is often used to fit $T_0$.

\begin{figure}[ht!]
\centering
	\includegraphics[width=\textwidth]{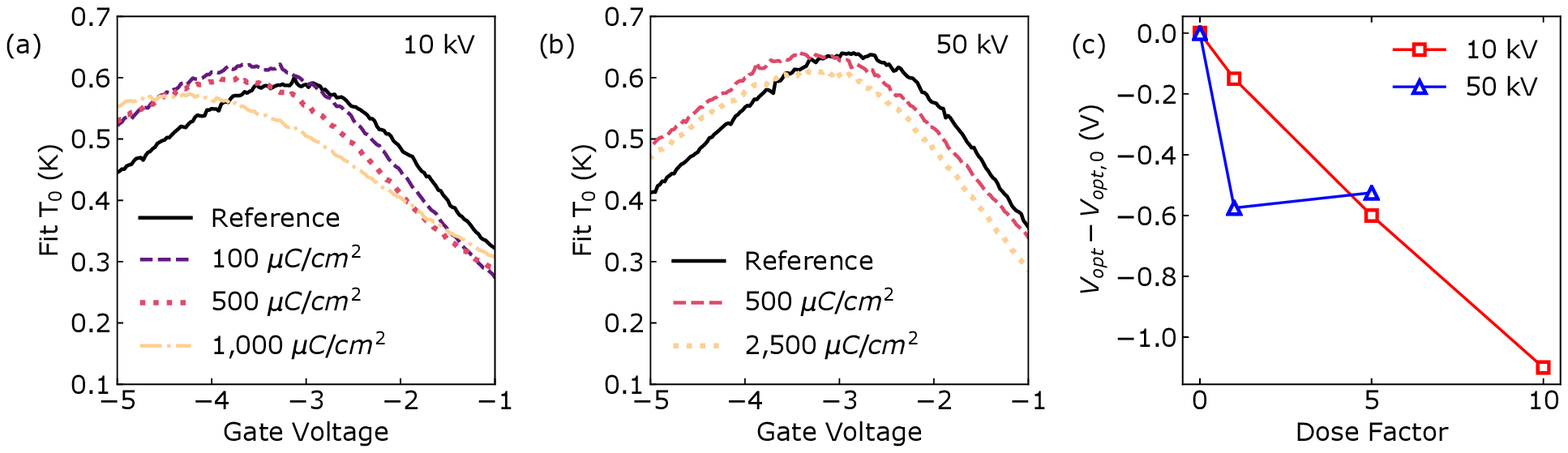}
	\caption{Temperature scales extracted from an Arrhenius plus offset model as a function of gate voltage for regions of Cr-BST Hall bars exposed at (a) 10 kV and (b) 50 kV. Maxima are taken as the effective size of the magnetic exchange gap, and the corresponding gate voltage as tuning the Fermi level to the center of the gap. (c)~Shifts in optimal gate voltage compared to reference regions for exposures at 10 kV (red squares) and 50 kV (blue triangles). Data is plotted as a function of dose factor relative to the clearing dose; at 10 (50) kV, the clearing dose is 100 (500) $\mu$C/cm$^2$.}
	\label{sfig-arrhenius}
\end{figure}

Fit values for $T_0$ are maximal when the Fermi level sits in the middle of the magnetic exchange gap: if the Fermi level is offset from the middle of the gap, the barrier to thermal excitation across the gap, and therefore the fit value of $T_0$, is reduced. Figure~\ref{sfig-arrhenius}(a,b) plots the temperature scale for thermally activated conduction for Hall bar regions exposed to electron beams under various conditions as a function of electrostatic gate voltage ($V_g$). The electrostatic gate voltage for which $T_0$ is maximal corresponds to the Fermi level sitting in the middle of the gap; this gate voltage is called the optimal gate voltage, $V_{opt}$. 

Shifts in optimal gate voltage between samples indicates a shift in the native Fermi level, caused by doping. Figure~\ref{sfig-arrhenius}(c) shows shifts in $V_{opt}$ as a function of dose factor for exposures at 10 and 50 kV. All data points are compared to the optimal gate voltage in the reference regions, $V_{opt,0}$. For the 10 kV region exposed at the clearing dose, $V_{opt} - V_{opt,0} < -0.2$~V. Comparing decreases in $T_0$ to the magnitude of gate voltage excursions, -0.2 V is considered a negligible shift in $V_{opt}$. This minimal shift in $V_{opt}$, which increases linearly with dose factor, indicates a systematic doping of the Cr-BST material with 10 kV electron beam exposure. For regions exposed at 50 kV, however, $V_{opt} - V_{opt,0}$ is a more substantial -0.6 V for the exposure at the clearing dose. Additionally, $V_{opt}$ does not shift monotonically with dose. 

\begin{figure}[ht!]
\centering
	\includegraphics[width=0.8\textwidth]{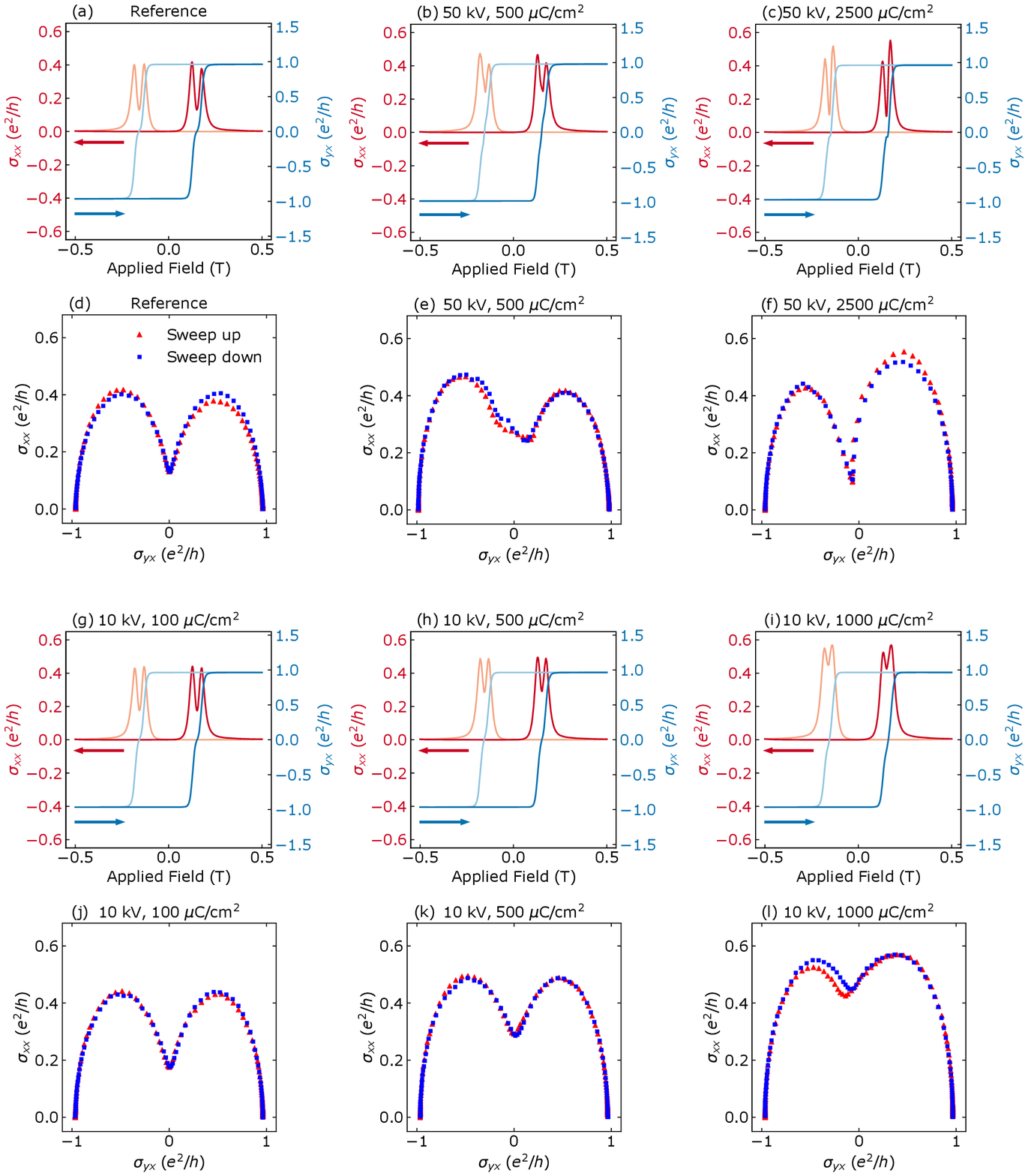}
	\caption{Extended dataset for electronic transport measurements of Cr-BST Hall bars across magnetization reversal. Vertical pairs of subplots correspond to measurements of a specific localized region that was exposed with an electron beam as denoted in the subplot title. Data in subplots (a, b, d, f, g) are reprinted from Figure~\ref{fig2}. (a-c,g-i) Longitudinal and Hall conductivities as a function of applied magnetic field. Longitudinal conductivity is shown in red and plotted on the left axis, while Hall conductivity is shown in blue and plotted on the right axis (axis indicated by arrows). Data taken while sweeping field from negative to positive (positive to negative) is shown in darker (lighter) colors. (d-f,j-l) Parametric plots of longitudinal conductivity as a function of Hall conductivity as magnetic field is increased (red triangles) or decreased (blue squares). All data was taken at or near the optimal gate voltage for each pair: (a,d) $V_g = -3$~V. (b,c,e,f) $V_g = -3.2$~V. (g,h,j,k) $V_g = -3.8$~V. (i,l) $V_g = -4.2$~V.}
	\label{sfig-qah-ext}
\end{figure}

Figure~\ref{sfig-qah-ext} shows measurements of longitudinal and Hall conductivity acquired as a function of applied magnetic field for all of the Cr-BST Hall bar regions shown in Figure~\ref{sfig-QAH-pics}(a) as well as the regions exposed at 50 kV shown in Figure~\ref{sfig-QAH-pics}(b). A subset of this dataset is shown in Figure~\ref{fig2}. Regions of the Hall bar shown in Figure~\ref{sfig-QAH-pics}(b) that duplicate regions shown in Figure~\ref{sfig-QAH-pics}(a) produced quantitatively similar measurements to their counterparts. 

Regions exposed at 10 kV deviate monotonically from the behavior shown in the reference region. All data shown in Figure~\ref{sfig-qah-ext} were taken at or near  $V_{opt}$ for each region, which rules out Fermi level shifts as the cause of this variation. Rather, it appears that behavior across the topological phase transition changes with dose after 10 kV~exposures. In the reference region and the region exposed at the 100 $\mu$C/cm$^2$ clearing dose, the system tunes somewhat explicitly through a trivial insulating $C=0$ phase at magnetization reversal as discussed in the main text. As dose is increased, the system appears to tend towards a direct $C=-1 \leftrightarrow C=+1$ transition. 

Regions exposed at 50 kV, however, demonstrate different behavior. A repeatable, non-hysteretic asymmetry in behavior across magnetization reversal is apparent even at the 500 $\mu$C/cm$^2$ dose. While we currently lack an explanation for this dramatic change in phenomenology, we posit it may indicate higher damage at one surface of the Cr-BST thin film than the other. Since the film under study is a modulation-doped film with an enhanced Cr concentration at the top and bottom surface, selective damage to one surface over the other may introduce some asymmetry upon reversing magnetization.

The differences in behavior between regions exposed at 10 versus 50 kV suggest an accelerating-voltage-dependent damage mechanism even when controlling for dose; comparing regions exposed with a 500 $\mu$C/cm$^2$ dose, the region exposed at 10 kV performs more similarly to the reference region than the region exposed at 50 kV.

\section{Additional test lithography data}

\begin{figure}[ht!]
\centering
	\includegraphics[width=\textwidth]{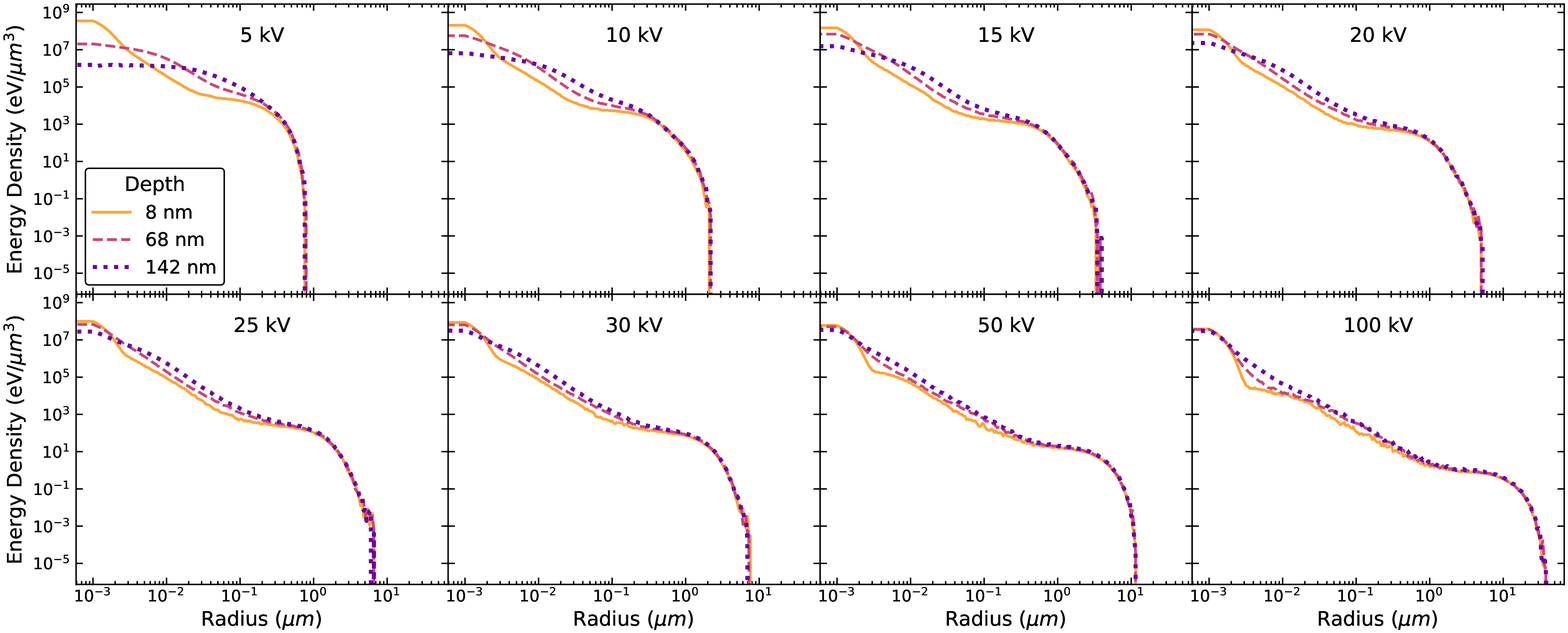}
	\caption{Simulated point spread functions for an electron beam exposure on a 150 nm PMMA / 8 nm \ce{Sb2Te3} / 0.5 mm GaAs stack for accelerating voltages 5-100 kV. Each subplot shows linecuts of energy density as a function of radius from the center of the electron beam point source at several depths into the PMMA in the same format as Figure~\ref{fig3}(a,b). Linecuts at 8 nm, close to the top of the PMMA, are shown as yellow solid lines. Linecuts at 68 nm, towards the middle of the PMMA, are shown as pink dashed lines. Linecuts at 142 nm, close to the surface of the substrate, are shown as purple dotted lines. }
	\label{sfig-extra-tracer}
\end{figure}

Figure~\ref{sfig-extra-tracer} shows point spread functions calculated in the same manner as those shown in Figure~\ref{fig3}(a,b) for a range of acccelerating voltages 5-100~kV. From 100~kV down to 25-30~kV, the bulk of the energy density is deposited within 10 nm of the beam center. For lower accelerating voltages, small-angle forward scattering becomes more significant and the energy density spreads laterally as electrons progress down into the resist stack. The low plateaus in energy density at radii far from the beam center are caused by electrons that backscatter off of the \ce{Sb2Te3} surface and return upwards through the resist.  

\begin{figure}[ht!]
\centering
	\includegraphics[width=0.4\textwidth]{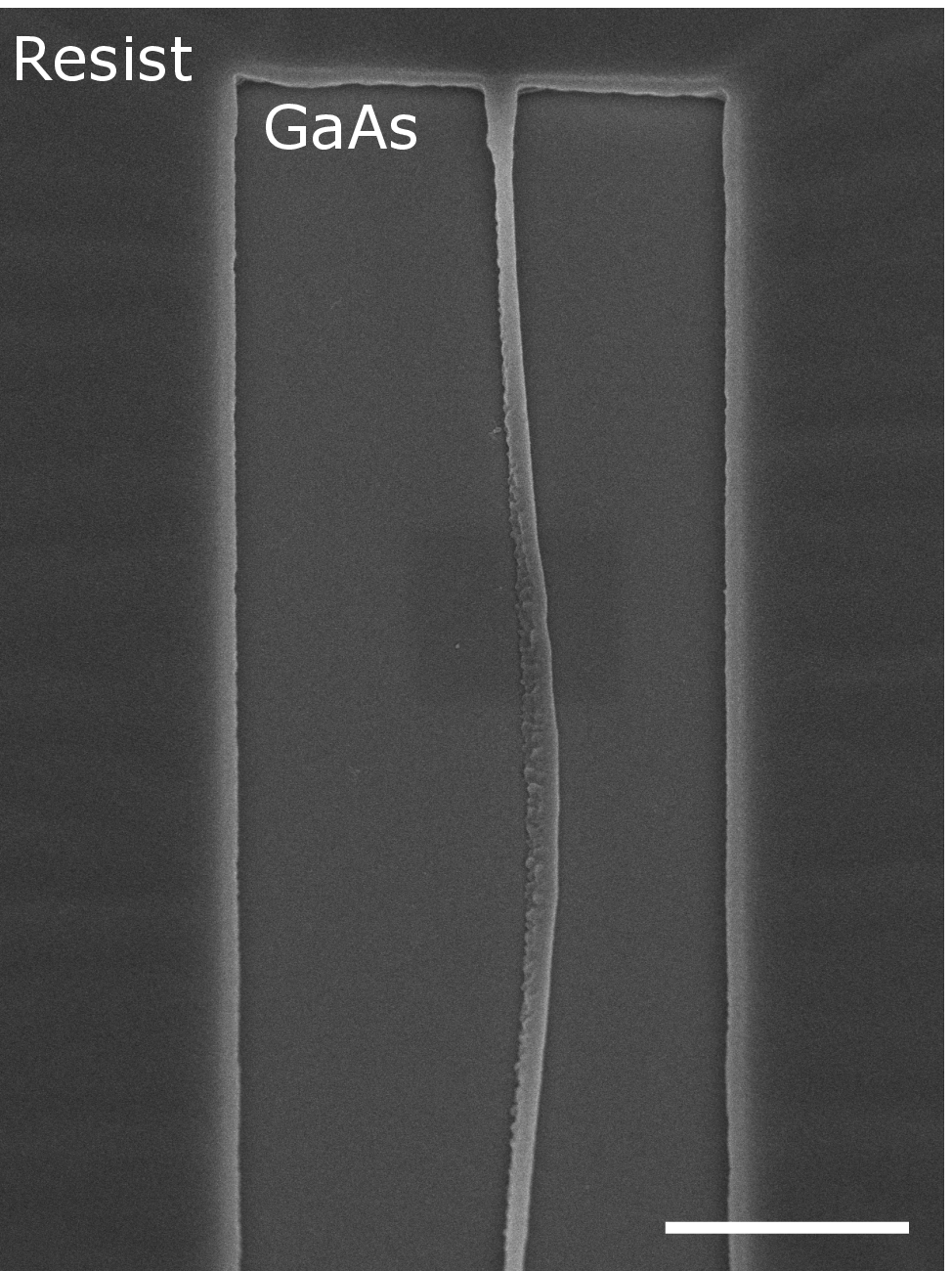}
	\caption{SEM image of developed resist after exposure with a 10 kV electron beam. Because the resist was too thick, the thin resist bridge was completely undercut and delaminated from the substrate. The slightly darker square in the middle of the image is a result of carbon contamination deposited during acquisition of a different SEM image. Scale bar: $1 \mu m$}
	\label{sfig-delamination}
\end{figure}

The risk associated with small-angle scattering of electrons through thicker resists is demonstrated in Figure~\ref{sfig-delamination}, which shows a thin resist bridge that completely delaminated from the substrate after patterning at 10 kV and development. In this case, A5 950 PMMA was spin-coated onto a bare GaAs substrate and baked for 5 minutes at 80\degree~C. The resist was exposed with a $100 \mu$C/cm$^2$ dose and a 10 kV accelerating voltage on a Raith VOYAGER electron beam lithography system and developed for 55/20 s 1:3 MIBK:IPA/IPA. A 4 nm 60/40 Au/Pd charge-neutralizing layer was added before imaging. 

\begin{figure}[ht!]
\centering
	\includegraphics[width=0.8\textwidth]{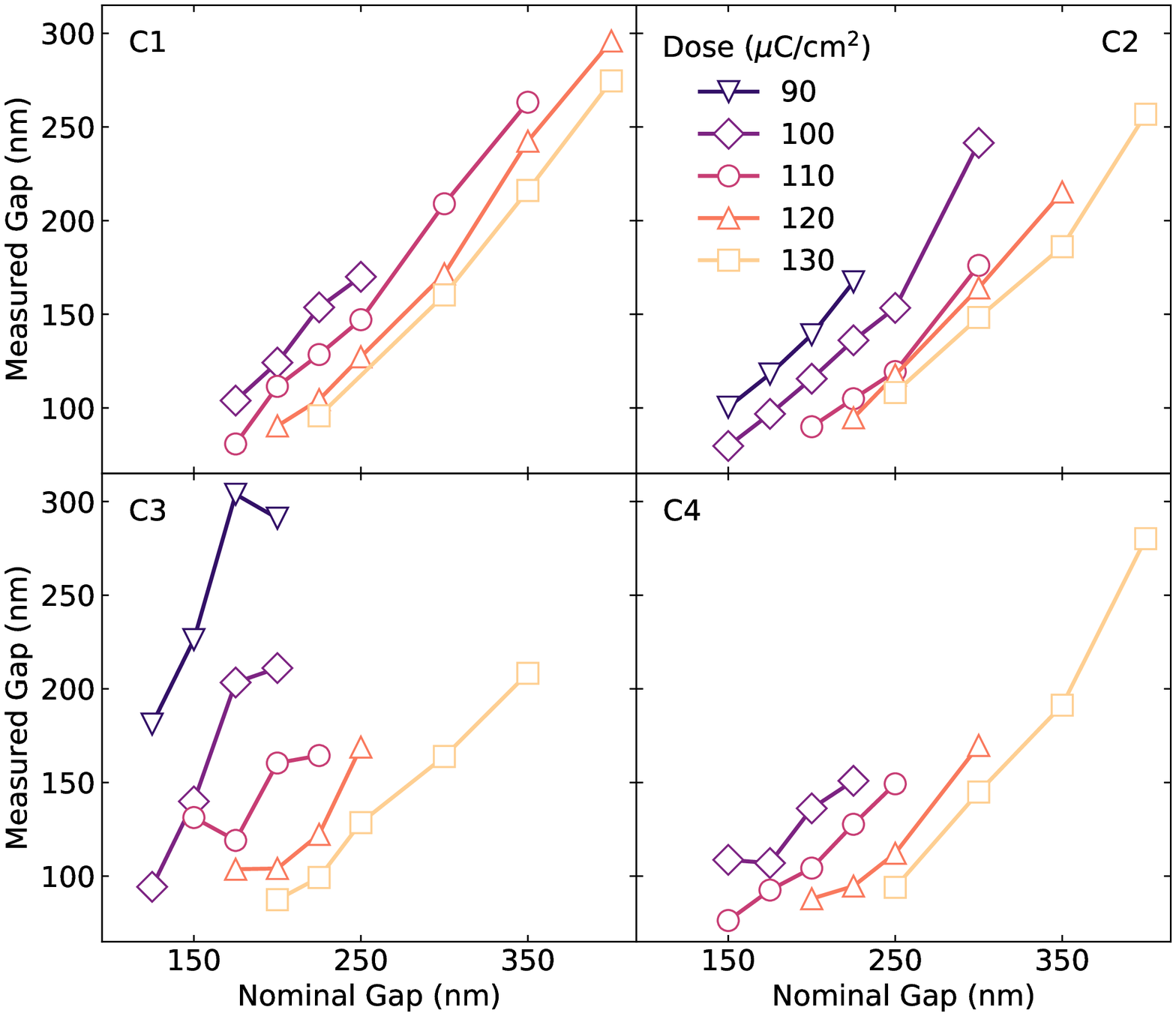}
	\caption{Full data for size of thin gaps patterned at 10 kV on C1-C4; processing details are described in the main text. Each subplot shows measured gap size after metallization and liftoff plotted against the nominal written gap size. Individual lines represent exposures at a specific dose. Where specific doses or nominal gap sizes are omitted, either liftoff failed and metal shorted across the gap or gaps were quite large. Data shown in Figure~\ref{fig4}(a) comes from the dose that produced the smallest gaps.}
	\label{sfig-TI-gaps}
\end{figure}

As described in the main text and shown in Figure~\ref{fig4}(a), EBL writes with a 10 kV accelerating voltage were used to produce thin gaps of BST after resist development, metallization, and liftoff. The complete dataset, including all e-beam doses and nominal gap sizes for all four chips C1-C4, is shown in Figure~\ref{sfig-TI-gaps}. For all four chips, measured gaps below 100 nm in width were attained for some set of patterning parameters.

Nominal gap sizes are necessarily larger than the measured gap sizes due to broadening of the pattern during the write. Where nominal gap sizes become too small, typically $<150$ nm, the thin resist bridge left behind after development can become completely undercut and delaminate entirely (Figure~\ref{sfig-delamination}). For chips C1 and C2, nominal gap size nearly linearly tunes the measured gap size below 150 nm. For C3 and C4, measured gaps changes monotonically but not quite linearly with nominal gap size.

\begin{figure}[H]
\centering
	\includegraphics[width=0.8\textwidth]{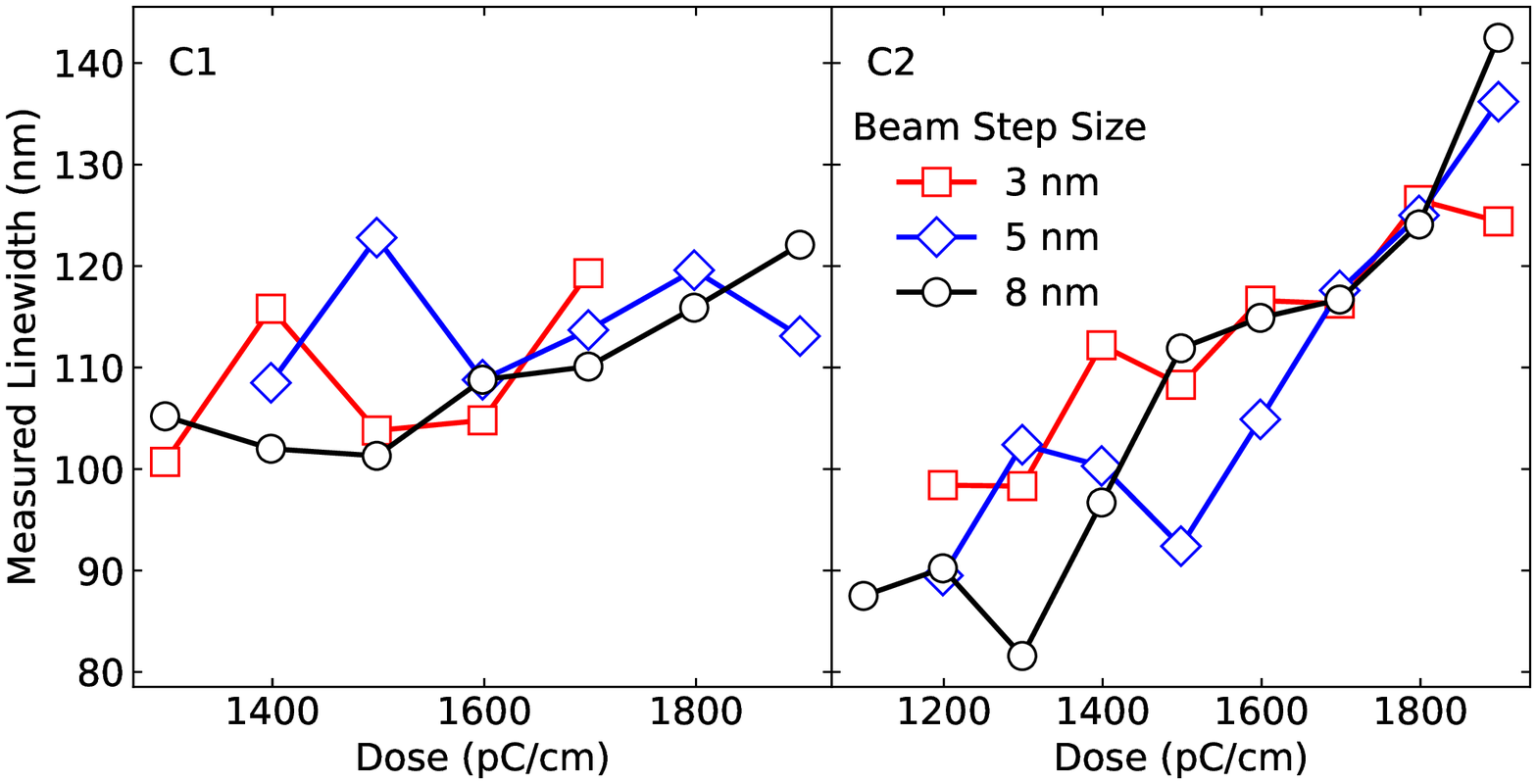}
	\caption{Full data for thin lines exposed at 10 kV on C1 and C2. Each subplot shows measured line width size after metallization and liftoff plotted against the nominal written gap size. Individual traces represent exposures with varying beam step sizes.}
	\label{sfig-TI-lines}
\end{figure}

Complete datasets for thin lines patterned with a 10 kV accelerating voltage are shown in Figure~\ref{sfig-TI-lines} for C1 and C2. In addition to single pixel lines written with a 3 nm beam step size (as described in the main text and shown in Figure~\ref{fig4}(b)), SPLs were also written with 5 and 8 nm step sizes. For all three step sizes, lines thinner than 120 nm were consistently produced across a wide range of doses.  

On C1, linewidths remain consistently 100-120 nm across all measured doses. On C2, in contrast, dose roughly monotonically tunes the measured linewidth between $\sim$90-140 nm over the same dose range.

\end{document}